\documentclass[%
 aip,jcp
 amsmath,amssymb,
 reprint,%
]{revtex4-1}

\usepackage{graphicx}
\usepackage{dcolumn}
\usepackage{amsmath}
\usepackage{bm}        
\usepackage{amssymb}   
\usepackage{xcolor}

\begin{document}

\preprint{AIP/123-QED}

\title{Generalized Voigt broadening due to thermal fluctuations of electromechanical nanosensors and molecular electronic junctions}

\author{Maicol A. Ochoa}
\affiliation{Biophysics Group, Microsystems and Nanotechnology Division, Physical Measurement Laboratory, National Institute of Standards and Technology, Gaithersburg, MD 20899}
\affiliation{Maryland Nanocenter, University of Maryland, College
Park, MD 20742}


\author{Michael Zwolak}
\email{mpz@nist.gov}
\affiliation{Biophysics Group, Microsystems and Nanotechnology Division, Physical Measurement Laboratory, National Institute of Standards and Technology, Gaithersburg, MD 20899}


\date{\today}

\begin{abstract}
  Graphene and other 2D materials give a platform for electromechanical sensing of biomolecules in aqueous, room temperature environments. The electronic current changes in response to mechanical deflection, indicating the presence of forces due to interactions with, e.g., molecular species. We develop illustrative models of these sensors in order to give explicit, compact expressions for the current and signal-to-noise ratio. Electromechanical structures have an electron transmission function that follows a generalized Voigt profile, with thermal fluctuations giving a Gaussian smearing analogous to thermal Doppler broadening in solution/gas-phase spectroscopic applications. The Lorentzian component of the profile comes from the contact to the electrodes. After providing an accurate approximate form of this profile, we calculate the mechanical susceptibility for a representative two-level bridge and the current fluctuations for electromechanical detection. These results give the underlying mechanics of electromechanical sensing in more complex scenarios, such as graphene deflectometry.
\end{abstract}

\maketitle

 Nanoscale sensing devices capable of operating in aqueous, ionic environments are highly desirable for selective molecular detection\cite{qi2003toward,sheehan2005detection,huang2008adsorption}, DNA sequencing\cite{zwolak2005electronic, lagerqvist2006fast, lagerqvist2007comment, lagerqvist2007influence, zwolak2008colloquium, chang2010electronic, huang2010identifying, tsutsui2010identifying, tsutsui2011electrical,tsutsui2012transverse} and cell biology studies\cite{dias2002electrochemical}. Due to their unique electromechanical properties, graphene and other carbon-based nanostructures are ideal active components in nanoelectromechanical switches\cite{shi2012studies,zhang2016electro} and nanoscale sensors\cite{traversi2013detecting,hierold2007nano,bruot2015tuning,boland2016sensitive,gruss2017communication,gruss2018graphene,heerema2018probing}. The design of detection protocols requires a quantitative correspondence between electron transport and mechanical deflection. In hot, wet environments, sensitive detection must also account for the effect of thermal fluctuations and other sources of noise. These effects are imprinted on the average electronic transmission function, which depends parametrically on the structural and electronic properties of the active sensing component, which is responsive to environmental perturbations.   

 Transport properties in molecular electronic junctions are also sensitive to inhomogeneities and thermal fluctuations. Molecules in junctions and active materials in nanoscale electromechanical sensors, follow similar principles in their transport properties, and their response to environmental perturbations is comparable. Due to the heterogeneity in the structural configuration of the molecule and electrode geometry in the formation of molecular junctions, conductance measurements yield different results in similar systems and follow a distribution, recorded in the form of histograms\cite{xu2003measurement,venkataraman2006dependence}. The peaks in these distributions are ascribed to molecular conductance channels, while the spread results from different sources, such as changes in tunneling length, substrate roughness, tip chemistry, presence of solvent, and extensive tip usage\cite{baldea2012interpretation, williams2013level, quan2015quantitative}. In other words, on the whole, the conductance distribution is not due to direct thermal fluctuations of the constituents of the molecule, which happen at a more rapid timescale and are averaged over in the measurements. However, in some proposed sensors, such as suspended graphene ribbons\cite{gruss2017communication,gruss2018graphene}, the fluctuations can be very slow -- on the order of  nanoseconds-- and commensurate with sensing. Moreover, they are well-separated from the timescale for electron transport, allowing for a Born-Oppenheimer treatment of their influence, which is the approach we will take here.

 We calculate the electronic transmission function for representative model systems including the effect of mechanical fluctuations of thermal origin. This permits us to characterize the signal-to-noise ratio for electromechanical sensing in nanoscale deflectometers, as well as understand thermally induced broadening. Studies of this kind, where different broadening mechanisms affect the output signal, are common in spectroscopy, where Doppler effects transform the Lorentzian lineshape into a different distribution: the Voigt profile\cite{quan2015quantitative,gruss2018graphene}. In the context of nanoscale electronics, thermal fluctuations of a single electronic level coupled to two fermionic baths transform the transmission function into exactly the Voigt profile. In the regime where thermal fluctuations dominate the noise, this profile for the transmission function is a linear combination of a Gaussian and an error function\cite{whiting1968empirical,gruss2018graphene}. For models beyond a single level, the transmission function is a generalized Voigt profile, i.e., a Gaussian “bulk” with more complex, algebraic tails. For these systems, we identify changes in the stationary current as a function of mechanical deflection, as well as characterize the linear response of the system in terms of the electromechanical susceptibility.  This simplified setting gives the underlying mechanics to electromechanical detection of molecular forces and structural fluctuations. Moreover, they provide new insights into the study of mechanical stress and thermal contributions to molecular electronics \cite{franco2011tunneling,inatomi2015effect,koch2018structural}.


{\bf The Voigt Profile.} Consider a single energy level $\varepsilon_p$ coupled to two metallic contacts. When each metal acts approximately as a noninteracting fermionic wide-band limit reservoir, the resulting coupling strength $w$ is energy independent and the level energy broadening is given by a Lorentzian distribution. Beyond the wide-band limit, the coupling strength depends on the energy $\varepsilon$ according to the spectral function of the reservoir and the energy brodeaning may follow other distributions. For the present discussion, we restrict ourselves to the case in which the wide-band limit is a reasonable approximation. In addition, we take the same coupling strength $w$ for each contact. The transmission function $T(\varepsilon)$ is therefore
\begin{align}
  \label{eq:T_one}
T(\varepsilon - \varepsilon_p) =& w A(\varepsilon - \varepsilon_p),
\end{align}
with 
\begin{align}
  A(\varepsilon - \varepsilon_p) =& \frac{w}{(\varepsilon - \varepsilon_p)^2+w^2}.
\end{align}
Thermal fluctuations introduce additional (inhomogeneous) broadening to the energy level. When these are due to many independent sources, the energy level will follow a Gaussian distribution $g(\varepsilon_p)$ centered at some equilibrium value $\bar \varepsilon_p$
\begin{align}
  \label{eq:G_one}
g(\varepsilon_p) =& \frac{1}{\sqrt{2 \pi \sigma^2}} e^{-\frac{(\varepsilon_p - \bar \varepsilon_p)^2}{2 \sigma^2}},
\end{align}
with standard deviation $\sigma$. In terms of the transmission function $T$, the stationary electronic current, which averages over all thermodynamic fluctuations, is given by 
\begin{align}\label{eq:current}
  \langle I \rangle  = \frac{2 e}{h}\int d\varepsilon \langle  T (\varepsilon) \rangle [f_{\mathcal{L}}(\varepsilon) - f_{\mathcal{R}}(\varepsilon)],
\end{align}
where $f_{\mathcal{L}/\mathcal{R}}(\varepsilon)=(\exp[\beta (\varepsilon-\mu_{\mathcal{L}/\mathcal{R}})]+1)^{-1}$ is the Fermi distribution function, $\beta$ is the inverse temperature in units of energy, $\mu_{\mathcal{L}(\mathcal{R})}$ is the chemical potential in the left (right) reservoir and $\langle T(\varepsilon) \rangle$ is the thermal average of the transmission function given by 
\begin{equation}
  \label{eq:Voigt_one}
  \langle T(\varepsilon) \rangle = \int d\varepsilon_p T(\varepsilon - \varepsilon_p) g(\varepsilon_p) = T*g.
\end{equation}
In Eq.\ \eqref{eq:current}, the factor of 2 accounts for the spin. The symbol $*$ in Eq.\ \eqref{eq:Voigt_one} represents the convolution operation. The expression in Eq.\ \eqref{eq:Voigt_one} is proportional to the Voigt profile  $V(\varepsilon) = A(\varepsilon) * g(\varepsilon)$ frequently found in molecular spectroscopy and diffraction studies\cite{wertheim1974determination,boone2007speed}, and takes on the form (see Appendix A) 
\begin{align}
\langle T (\varepsilon) \rangle =& w {\rm Re}\left[\left(\sqrt{\frac{\pi}{2 \sigma^2}} - J(E, \sigma) \right) e^{\frac{E^2}{2 \sigma^2}}\right],   \label{eq:Voigt_one_avg}
\intertext{ with $E = i(\varepsilon - \bar \varepsilon_p) + w$ and}
  J(E, \sigma) =& \frac{E}{\sigma^2} \int_0^1 d \alpha e^{-\frac{\alpha^2 E^2}{2 \sigma^2}} = \sqrt{\frac{\pi}{2}}{\rm erf} \left( \frac{E}{\sqrt{2}\sigma} \right). \label{eq:Jfunc}
 \end{align}\\

\begin{center}
  \begin{figure}[t]
    \centering
    \includegraphics[scale=0.4]{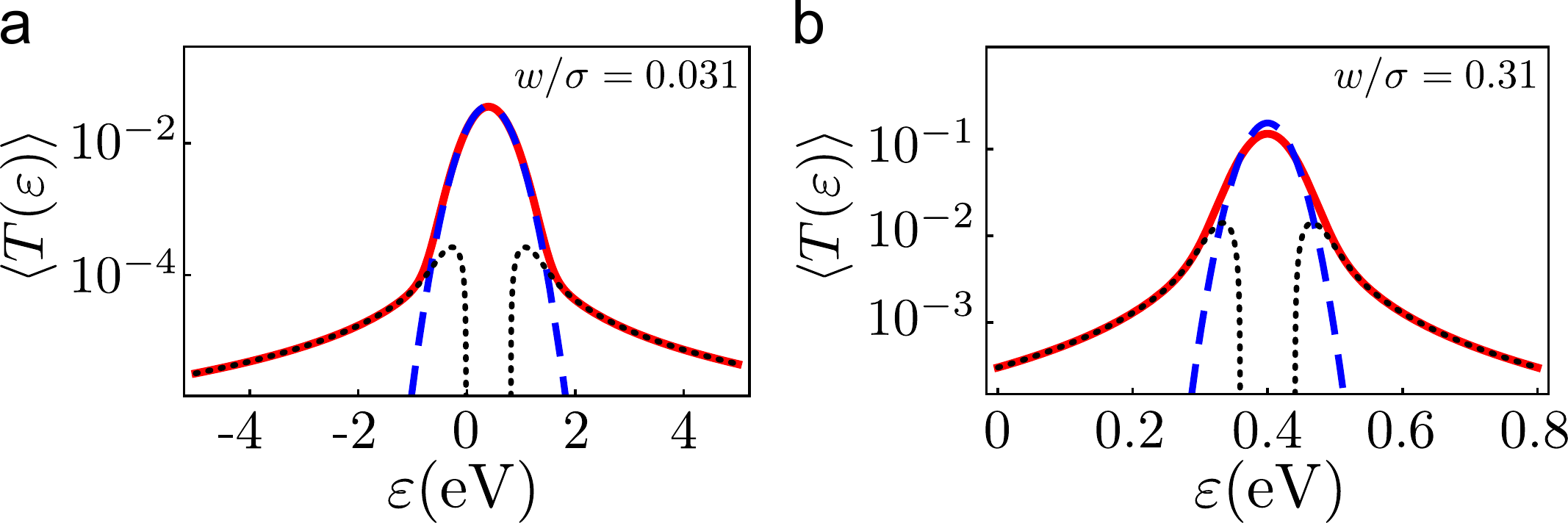} 
    \caption{(Color online) Voigt profile for the thermally averaged transmission function for a single level in contact with two reservoirs. (a,b) The Voigt profile calculated from numerical convolution (red, solid), the Gaussian component of $\langle T \rangle$ (blue, dashed), and the algebraic tail (black, dotted) of $\langle T \rangle $. Parameters for this model are $\bar \varepsilon_p= 0.4$ eV and $w= 0.01$ eV. In (a), the thermally induced broadening -- which is proportional to $1/\beta$ -- is $\sigma^2 =0.1$~eV$^2$ and, in (b), $\sigma^2= 0.001$~eV$^2$. When $\sigma > w$, there is a clear separation of the Gaussian and algebraic components in Eq.\ \eqref{eq:Voigt_one_avg}. The error bars in the numerical integration are smaller than the line width.}
    \label{fig:ols}
  \end{figure}
\end{center}

\vspace{-1.2cm}
Equation \eqref{eq:Voigt_one_avg} is along the lines of previous expressions for the Voigt profile in terms of the error function ({\rm erf}) and the Faddeeva function\cite{humlivcek1982optimized,olver2010nist}. In Fig.\ \ref{fig:ols}, we compare the transmission function obtained from numerical integration of Eq.\ \eqref{eq:Voigt_one} and the compact form, Eq.\ \eqref{eq:Voigt_one_avg}, which separates the Gaussian bulk of the peak from the algebraic tails. Utilizing the factorial series for the error distribution\cite{olver2010nist} and Eq.\ \eqref{eq:Jfunc}, one can see that the leading term far from the peak comes from the error function and is proportional to ${\rm Re}\{1/E\}$, i.e., the Lorentzian tail.

 The above takes a Born-Oppenheimer treatment by employing an uncorrelated classical random variable.  This can be extended to account for correlations. In the Supplemental Information (SI), we examine non-Markovian effects between a local vibration and the electronic site. This gives rise to an additional energy-dependent correction on top of the Voigt profile. Inelastic effects -- the emission/absorption of vibrational quanta by transporting electrons\cite{klein1973inelastic,segal2002conduction, chen2003local, chen2004inelastic,yang2005role,chen2005inelastic,galperin2007molecular,reed2008inelastic} -- can also be included. However, for electromechanical sensors in solution at room temperature, such as the graphene deflectometer\cite{gruss2018graphene}, the relevant vibrational modes have small energies compared to room temperature and are noisy due to the solution environment. Thus, coherent effects should not significantly contribute.


\vspace{-1.0cm}


\begin{center}
  \begin{figure}[b]
    \centering
    \includegraphics[scale=0.6]{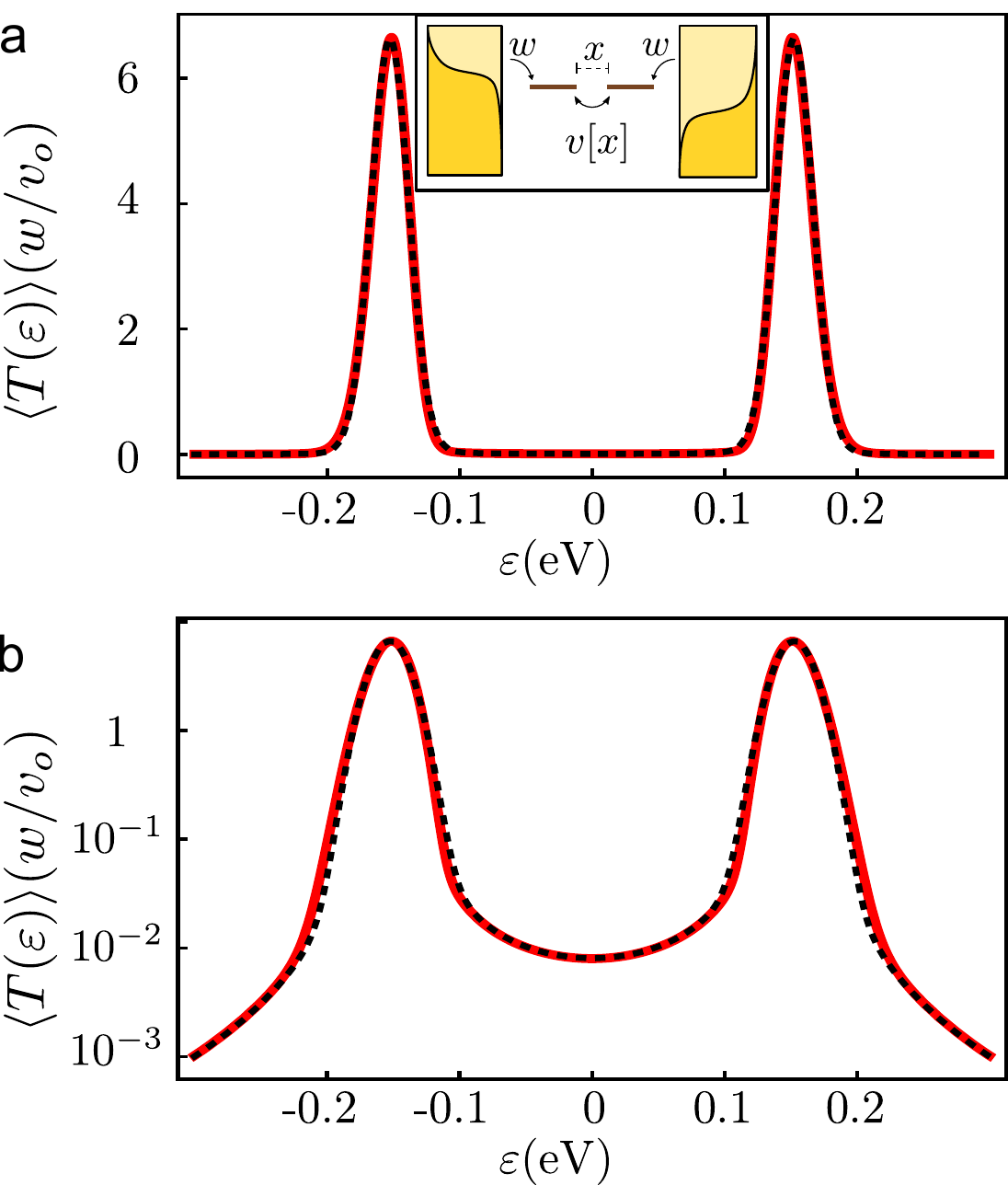} 
    \caption{(Color online) Thermal broadening of the electron transmission function for a two-level system.  (a,b) $\langle T \rangle$ in linear and logarithmic scale obtained directly from numerical convolution (red, solid), and from the analytic result in Eq.\ \eqref{eq:T2_full} (black, dotted). Inset: schematic representation of the system. Two identical energy levels localized at different positions, at a relative distance $x$, are coupled to two leads with the same coupling constant $w$. The electron tunneling strength $v$ between the two levels depends parametrically on $x$ according to Eq.\ \eqref{eq:tunnel}. Thermal fluctuations and local forces modify $x$, affecting the transport properties of the system.  Parameters are such that they reproduce the first transmission peak and its electromechanical response for suspended graphene nanoribbons\cite{gruss2018graphene}:  $\lambda = 0.047$ nm, $\bar x = 0.14$~nm, $w = 1.3$~meV, $\kappa = 1400$~eV/nm$^2$, $v_o = 0.153$~eV, $\bar \varepsilon_p = 0$, $\mu = 0$, and at $300$ ~K . The Lorentzian broadening $w$ is such that $w \ll \sigma$ holds for this system.}  
    \label{fig:tls}
  \end{figure}
\end{center}


{\bf Generalized Voigt profile}. The electronic structure of nanoscale systems, such as molecules and 2D materials, are often taken as tight-binding models.  In such models, the system $\mathcal{S}$ is composed of noninteracting electronic states with Hamiltonian
 \begin{align}
 H_{\mathcal{S}} =\sum_{i \in \mathcal{S}} \varepsilon_i \hat c_i^\dagger \hat c_i + \sum_{i\neq j} v_{ij} \hat c_i^\dagger \hat c_j, 
 \end{align}
where $\hat c_i$ ($\hat c_i^\dagger$) is the annihilation (creation) operator for an electron in the $i^{\rm th}$ state, with corresponding single electron energy $\varepsilon_i$ and tunneling constant $v_{ij}$.  The electronic structure of each metallic region, on the left $(\mathcal{L})$ and right $(\mathcal{R})$, is a fermionic reservoir with Hamiltonian $H_{\mathcal{L}/\mathcal{R}} = \sum_{k \in \mathcal{L}/\mathcal{R}} \varepsilon_k \hat c_k^\dagger \hat c_k$. Here,  $\varepsilon_k$ is the corresponding single particle energy for an electron in state $k$ of the reservoir, and $\hat c_k$ ($\hat c_k ^\dagger$) is the corresponding annihilation (creation) operator. The coupling between the system and the reservoirs is given by the bilinear form  
\begin{align}\label{eq:HI}
  H_{\mathcal{I}} = \! \sum_{k \in \mathcal{L}, \mathcal{R}, i \in \mathcal{S}} \! v_{ik} \hat c_i^\dagger \hat c_k +  v_{ik}^* \hat c_k^\dagger \hat c_i,
\end{align}
where $v_{ik}$ is the coupling between the $i^{\rm th}$ orbital in $\mathcal{S}$ and  $k^{\rm th}$ state in $\mathcal{L}, \mathcal{R}$.  
 
The stationary current originating from an applied bias $eV = \mu_{\mathcal{L}} - \mu_{\mathcal{R}}$  has the form of  Eq.\ \eqref{eq:current}  in terms of the thermally broadened transmission function  $\langle T(\varepsilon) \rangle = T * g$ with
\begin{align}\label{eq:TN}
    T (\varepsilon) =&  {\rm Tr}\{\Gamma_R(\varepsilon) G^r(\varepsilon) \Gamma_L(\varepsilon) G^a(\varepsilon) \},
\end{align}
where ${\rm Tr}$ is the trace and $g$ is the thermal distribution function. In Eq.\ \eqref{eq:TN}, the retarded (advanced) Green function $G^r$ ($G^a$) for the system and the coupling matrices $\Gamma_{\mathcal{L}/\mathcal{R}}$, are given by $[\Gamma_{\mathcal{L}/\mathcal{R}}]_{ij}~=~2\pi~\sum_{k \in \mathcal{L}/\mathcal{R}}~v_{k,i} v_{k, j}^* \delta(\varepsilon - \varepsilon_k)$, 
\begin{align}
  G^r(\varepsilon) =& [\varepsilon \mathcal{I} - H_{\mathcal{S}} +(i/2)(\Gamma_{\mathcal{L}}(\varepsilon) + \Gamma_{\mathcal{R}}(\varepsilon)) ]^{-1},
\end{align}
and $G^a(\varepsilon) = G^r(\varepsilon)^\dagger$, with $\mathcal{I}$ the identity matrix. 

Mechanical and thermal fluctuations modify the interatomic distances $x_{ij}$ in the active region, rendering them with stochastic dynamics and a statistical distribution $g$ around equilibrium values $\bar x_{ij}$. These, in turn, influence the tunneling constants $v_{ij}$ according to 
\begin{align}
  v_{ij} = v_{ij}^o e^{-(x_{ij}-\bar x_{ij})/\lambda_{ij}}, \label{eq:tunnel}
\end{align}
where $\lambda_{ij}$ is the characteristic electronic decay length\footnote{The real coupling response will be more complicated. We could also include the effect of electron-vibration couplings in the model in the form illustrated in the SI. However, the contributions to the total signal originating from inelastic effects are small in the case of graphene-based nanosensors operating at room temperature. See Refs.\ \citenum{klein1973inelastic} and \citenum{reed2008inelastic} and Fig.\ S7 in the SI.}$^,$\cite{cosma2014strain}.  Importantly, $\bar x_{ij}$, and consequently $v_{ij}$, are responsive to local forces.

  We can gain further insight into the effects on the transmission function due to thermal fluctuations and local stress by considering the two-level bridge in the inset in Fig.\ \ref{fig:tls}a.  For this model, setting $v_{12} \equiv v$, $v_{ij}^o = v_o$, the transmission function is 
 \begin{align}
   T(\varepsilon) =&\frac{w^2 v^2}{[(\varepsilon - iw/2)^2-v^2][(\varepsilon + iw/2)^2-v^2]}.\label{eq:lorentzian}
 \end{align}
 The interparticle distance $x$ influences $v$ and is subject to thermal fluctuations around its equilibrium value $\bar x$. As a result, the transmission function fluctuates in time, and its average $\langle T(\varepsilon) \rangle$ over the distribution of configurations $g(x)$ is given by the convolution integral $T*g$. For weak interparticle vibrations,  we can take this as harmonic  -- with characteristic force constant $\kappa$, and minimum at $\bar x$ -- such that $g(x)$ is normally distributed as
\begin{align}\label{eq:gij}
  g (x) =& \frac{1}{\sqrt{2 \pi \sigma^2}} e^{-\frac{(x -\bar x)^2}{2 \sigma^2}},
\end{align}
with variance given by thermal fluctuations $\sigma^2 = (\beta \kappa)^{-1}$.  The equilibrium interparticle distance $\bar x$ varies as $\bar x \to  \bar x + F/\kappa$  when a local force, $F$, is present.

 When the characteristic decay length $\lambda$ is larger than the standard deviation for the interparticle distance fluctuation (i.e., $\lambda > \sigma$), the thermally broadened transmission function is well approximated by (see Appendix \ref{ap:T_avg_one} for details)
\begin{align}
  \langle T (\varepsilon)\rangle \approx& w^2 \sum_{k \in \{-1,1\}} {\rm Re}\left[ C \left(\sqrt{\frac{ \pi}{ 2\bar \sigma^2}} +  J(E_k, \bar \sigma) \right) e^{\frac{E_k^2}{2 \bar \sigma^2}}\right]. \label{eq:T2_full}
\end{align}
  In Eq.\ \eqref{eq:T2_full}, $C = (1/8) (i\varepsilon^{-1} + 2 w^{-1})$,  $E_k = i(\varepsilon +k v_o)-w/2$, and the standard deviation $\bar \sigma = \sigma v_o / \lambda$. Equation \eqref{eq:T2_full} is similar to Eq.\ \eqref{eq:Voigt_one_avg} but accounts for the explicit effect of mechanical fluctuations in the transmission function. This result indicates that when thermal fluctuations are the main broadening mechanism ($w < \sigma$), the variance in the observed distribution $\bar \sigma ^2$ is proportional to the temperature and inversely proportional to the square of the the characteristic decay length $\lambda$. In Figs.\ \ref{fig:tls}a,b,  we compare Eq.\ \eqref{eq:T2_full} to the exact numerical convolution $T * g$ for parameters representative of suspended graphene nanoribbons\footnote{We choose parameters to match the response and position of the first energy level to suspended nanoribbons. However, the actual values, e.g., for the mechanical coupling constants and other parameters will be different in some cases due to the difference in the setup (two sites versus a whole ribbon, etc.).}. This illustrates that the approximate analytic expression provides excellent agreement when $w < \sigma$. As in the case of the single level, the generalized Voigt profile in Eq.\ \eqref{eq:T2_full} has a Lorentzian decay for energies far from both peaks.

{\bf Sensing and current fluctuations}. Now we address the problem of detection of local forces upon measurement of stationary current in model electromechanical sensors. We search for conditions to maximize the difference in stationary current between the deflected and undeflected system $\Delta I$, as well as the signal-to-noise (SNR) ratio  
\begin{align}\label{eq:SNR_def}
  {\rm SNR} =& \frac{| \Delta I |}{\sigma_I},
\end{align}
where $\sigma_I$ is the standard deviation in the current.  

Fluctuations in the current $\delta I(t) = I(t) - \bar I$ originate from static and dynamical sources. For the model system with a single level described in Eqs.\ \eqref{eq:T_one}-\eqref{eq:Voigt_one_avg}, current fluctuations are due to the energy level fluctuations $\delta \varepsilon_p$, which are captured by the local distribution of energies $g(\varepsilon)$ in Eq.\ \eqref{eq:G_one}. Up to second order in $\delta \varepsilon_p$\footnote{Expanding around the current at the mean peak position $\bar{\varepsilon}_p$ up to third order,  $I(\varepsilon) \approx I(\bar \varepsilon_p)+ \delta \varepsilon_p\, \partial_{\varepsilon_p} I(\bar \varepsilon_p)+(\delta \varepsilon_p^2/2) \partial_{\varepsilon_p}^2 I(\bar \varepsilon_p)+(\delta \varepsilon_p^3/6)\partial_{\varepsilon_p}^3 I(\bar \varepsilon_p)$, and after noticing that $\left< \delta \varepsilon_p \right> = \left< \delta \varepsilon_p ^3 \right> =0  $, $\left< \delta \varepsilon_p ^4 \right> = 3 \sigma^4$ for the normal distribution in Eq.\ \eqref{eq:G_one},  the variance comes out to be Eq.\ \eqref{eq:sigmaI_one}.}, $\sigma_I^2 =\langle \delta I^2 \rangle$ takes the form
\begin{align}\label{eq:sigmaI_one}
\sigma_I^2 =& \left. \sigma^2  \left( \partial_{\varepsilon_p} I \right)^2 +\sigma^4 \left(\partial_{\varepsilon_p} I\, \partial_{\varepsilon_p}^3 I +\frac{1}{2}  \left( \partial_{\varepsilon_p}^2 I \right)^2\right) \right|_{\varepsilon_p = \bar \varepsilon_p}.
\end{align}
 
\noindent Before proceeding further, we make the remark that this approach may fail to describe the current statistics when large fluctuations are present, such as in the case of DNA sequencing with transverse transport where log-normal histrograms have been predicted\cite{zwolak2005electronic,lagerqvist2006fast, lagerqvist2007influence, lagerqvist2007comment,krems2009effect} and observed\cite{tsutsui2010identifying}. In this case, the relevant fluctuations are in the contact between the molecules and the electrodes, which are not covalently bonded.

 Local perturbations inducing a small but controlled shift in the energy level, $\bar \varepsilon_p \to \bar \varepsilon_p+ \Delta \varepsilon_p$, modify the stationary current as $I(\bar \varepsilon_p) \to I(\bar \varepsilon_p)+ \Delta I$ following the linear relation $\Delta I = \chi_\varepsilon \Delta \varepsilon_p$, with susceptibility 
\begin{align}\label{eq:suscep_one}
\chi_\varepsilon =& \frac{2 e }{h} \int d\varepsilon \left \langle \frac{d}{d \bar \varepsilon_p} T (\varepsilon) \right \rangle [f_{\mathcal{L}}(\varepsilon) - f_{\mathcal{R}}(\varepsilon)].
\end{align}
Significantly, to first order in the energy shift $\chi_\varepsilon = \partial_{\varepsilon_p} I|_{\varepsilon_p =\bar \varepsilon_p}$ and, therefore, the SNR in Eq.\ \eqref{eq:SNR_def} under small shifts in energy level takes the form 
\begin{align}\label{eq:SNR_one}
  {\rm SNR} \approx & \left. \frac{\left|\partial_{\varepsilon_p} I \Delta \varepsilon_p\right|}{\sigma \sqrt{ \left( \partial_{\varepsilon_p} I \right)^2 +\sigma^2 \left( \partial_{\varepsilon_p} I \, \partial_{\varepsilon_p}^3 I+\frac{1}{2} \left( \partial_{\varepsilon_p} ^2 I \right)^2\right)} } \right|_{\varepsilon_p =\bar \varepsilon_p}.
\end{align}
This approches $|\Delta \varepsilon|/ \sigma$ whenever $\sigma$ or $\partial_{\varepsilon_p}^2 I$ are small. As a consequence, in this limit the SNR improves with the energy shift $\Delta \varepsilon$ and worsens with energy level fluctuations originating from interactions with the environment.

We can extend the above analysis to the case of systems with mechanical deflections, such as the two-level bridge in Fig.\ \ref{fig:tls}. We consider the situation in which mechanical fluctuations dominate and the stochastic dynamics of the interparticle distance $x$ is described by, for example, a Langevin equation.  From this consideration, and for small perturbations in $\bar x$,  current and mechanical fluctuations are approximately related by

\begin{align}
\sigma_I^2 \approx &  \left. \frac{\left( \partial_x I \right)^2}{\beta \kappa}  + \frac{\partial_x I\,  \partial_x^3 I +\left(\partial_x^2 I \right)^2}{(\beta \kappa)^2} \right|_{x =\bar x} ,
\label{eq:x_variance}
\end{align}
with $\sigma^2 =(\beta \kappa)^{-1}$ as in Eq.\ \eqref{eq:gij}.

%
%


\begin{center}
  \begin{figure}[t]
    \centering
    \includegraphics[scale=0.435]{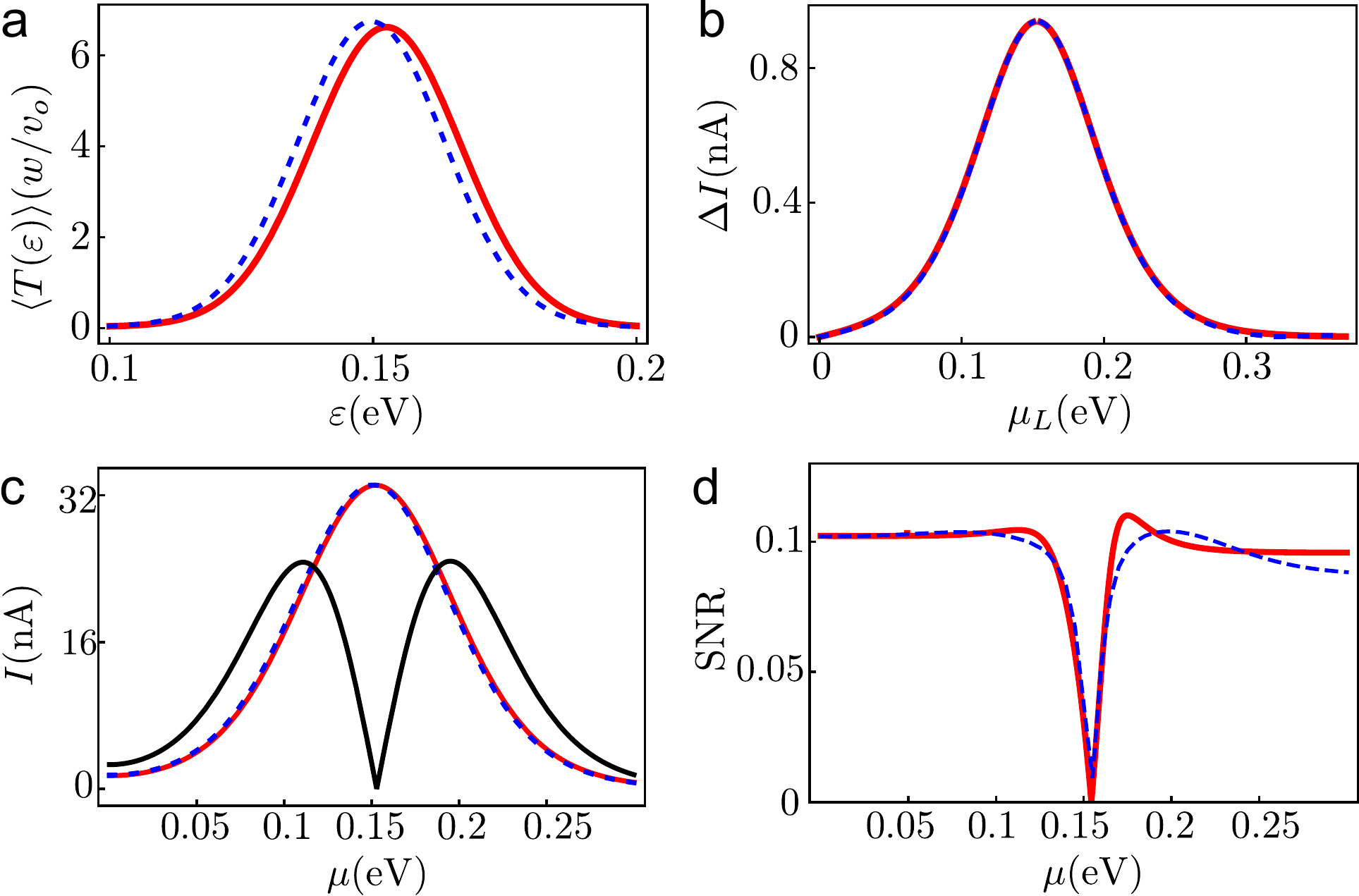} 
    \caption{(Color online) Transport and electromechanical sensing with the two-level system in Fig.\ \ref{fig:tls}a. (a) Energy shift in the transmission function induced by a local force: $F = 0$~pN  (red, solid) and $F = 100$~pN (blue, dashed). (b) Absolute difference in stationary current (red, solid) and utilizing Eq.\ \eqref{eq:T2_full} (blue, dashed) as a function of $\mu_{\mathcal{L}}$ ($\mu_{\mathcal{R}} = 0$). (c) Stationary current under symmetric bias $\mu_{\mathcal{L}} = \mu+0.025$~eV, $\mu_{\mathcal{R}} =\mu -0.025$~eV as a function of the Fermi energy $\mu$ and mechanical stress: $F = 0$~pN (red, solid) and $F = 100$~pN (blue, dashed). The absolute difference $|\Delta I|$ is also shown scaled by a factor of $40$ (solid, black), such that the $|\Delta I|_{\rm max} \approx 0.64$~nA. (d) Exact (blue, dashed) and approximate, Eq.\ \eqref{eq:SNR_1st_order} (solid, red), SNR for the sensing protocol in (c).  Other parameters are as in Fig.\ \ref{fig:tls}.}
    \label{fig:force}
  \end{figure}
\end{center}

\vspace{-1.05cm}

Next we consider the linear mechanical suceptibility of the sensor. In general, by considering linear deviations from equilibrium interatomic distances $\Delta x_{ij}$ in Eq.\ \eqref{eq:current}, we obtain the linear response in the stationary current in the form $\Delta I = \vec \chi \cdot \Delta \vec{x}$ with
\begin{align}\label{eq:chi}
  \vec \chi  =\frac{2 e }{h} \int d\varepsilon \langle \nabla T (\varepsilon) \rangle [f_{\mathcal{L}}(\varepsilon) - f_{\mathcal{R}}(\varepsilon)].
\end{align}

\noindent For the two-level system and to first order in the mechanical fluctuation $\langle \nabla T \rangle = \partial_x T(\bar x)$ and $\chi = \partial_x I(\bar x)$ and, consequently, the SNR is approximately given by
\begin{equation}
  \label{eq:SNR_1st_order}
  {\rm SNR } \approx \left. \frac{\sqrt{\kappa \beta} | \partial_x I \Delta x |}{\sqrt{\left(\partial_x I\right)^2+\frac{1}{\beta \kappa }\left(\partial_x I\,  \partial_x^3 I +\left(\partial_x^2 I \right)^2 /2\right)}} \right|_{x =\bar x} . 
\end{equation}
We notice that the SNR asymptotically approaches the limit $\sqrt{\kappa \beta} |\Delta x|$ when $\partial_x^2 I(\bar x)$ is small or when $\kappa$ is large. In such cases, it is clear that temperature deteriorates the SNR. Increasing the stiffness of the active material, though, improves the SNR. 

In Fig.\ \ref{fig:force} we illustrate the electromechanical properties of the two-level bridge studied in Fig.\ \ref{fig:tls}. We notice in Fig.\ \ref{fig:force}a that under mechanical deflection the thermally broadened transmission function shifts. This shift manifests in the absolute difference in the stationary current between the deflected and undeflected sensor (Fig.\ \ref{fig:force}b). By modulating the Fermi energy at a fixed symmetric bias, we find in Fig.\ \ref{fig:force}c a protocol for sensing local forces with maximal detection signal $\Delta I$ near Fermi energies corresponding to the maximum in the derivative of $I$ with respect to $\mu$. This result can be understood in light of the first order approximation to the susceptibility, $\chi = \partial_x I$. Indeed,  $\partial_x I = (d v/ dx ) \partial_v I$ and therefore the linear susceptibility is proportional to the derivative of the current with respect to the system energy, as the latter is determined by $v$.  Consequently, $\Delta I \sim \partial_\mu I$ whenever the modulation of $\mu$ affects the system energy in an equivalent manner to $v$. Figure \ref{fig:force}d compares the exact\footnote{We obtain numerically the exact SNR in three steps. First, we calculate $\Delta I = \left< I(\bar{x}) \right> -\left < I(\bar{x} + F/\kappa) \right > $ utilizing Eq.\ \eqref{eq:current}. Then, we compute the integral $\left< I ^{2} \right> = \int dx g(x) I(x) ^{2}$, where $I(x)$ is the instantaneous current for the interparticle distance $x$, given by  Eq.\ \eqref{eq:current} upon the substitution $\left< T \right> \to T(x)$. Finally, we take the ratio SNR $= |\Delta I|/ (\left< I ^{2} \right> -\left< I \right> ^2)^{1/2}$.} SNR with the approximate form in Eq.\ \eqref{eq:SNR_1st_order}, as a function of the Fermi energy and for the same protocol studied in Fig. \ref{fig:force}c. This result shows that Eq.\ \eqref{eq:SNR_1st_order} is a reliable approximation to the SNR at room temperature. Moreover, we find that the limit value of $\sqrt{\kappa \beta} |\Delta x| \approx 0.104$ is achieved at the tails of the current profile. The visible deviation is solely due to the fact that in the representative parameter regime, there are effects beyond linear response (see SI). The effect of finite sampling time, electrostatic fluctuations in models with a larger number of degrees of freedom that explicitly include mechanical fluctuations will be the subject of future investigations. 

In conclusion, we have obtained an analytic expression for the inhomegeneous broadening due to thermal fluctuation in the transmission function and the mechanical susceptibility. This generalized Voigt profile and the dynamical current fluctuations analyzed here provide a reliable mathematical description of the thermal contributions to the dispersion of conductivity measurements in electromechanical sensors and molecular junctions, whenever the mechanical deviations are small. Importantly, the generalized Voigt profile is the molecular electronics analog of the gas-phase spectroscopic lineshape. It thus allows one to understand measured currents and formulate protocols for electromechanical sensing at room temperature.\\

\vspace{-0.9cm}
\section*{Supplementary Material}
\vspace{-0.5cm}
See supplemental information for an extended analysis of our approximations to the current, fluctuations, and SNR, as well as a detailed study of the Voigt profile for a single level coupled to a local vibration.\\

\vspace{-0.9cm}
\begin{acknowledgments}
  \vspace{-0.5cm}
We thank Yonatan Dubi for helpful discussions. M. A. O. acknowledges support under the Cooperative Research Agreement between the University of Maryland and the National Institute of Standards and Technology Physical Measurement Laboratory, Award 70NANB14H209, through the University of Maryland. 
\end{acknowledgments}

\vspace{-0.55cm}
\appendix
\section{Derivation of equations Eq.\ \eqref{eq:Voigt_one_avg} and Eq. \eqref{eq:T2_full}} \label{ap:T_avg_one}
\vspace{-0.5cm}
We analytically integrate Eq.\ \eqref{eq:Voigt_one} by writing $T$ as a partial fraction decomposition $T =(w /2)\left( E^{-1} +E^{* \, -1}\right)$, with $E = i(\varepsilon - \varepsilon_p) +w$ and utilizing the identity
\begin{align}\label{eq:apEinv}
  \frac{1}{E} =& \int_0^\infty da\; e^{-a E},
\end{align}
which leads to the expression
\begin{align}
  \langle T \rangle =&  \int_0^\infty da \int d\varepsilon_p g(\varepsilon_p) \left( e^{-a E} + e^{-a E^*}\right).
\end{align}
The latter can be integrated, first with respect to $\varepsilon_p$ and then with respect to $a$ to give the result in Eq.\ \eqref{eq:Voigt_one_avg}. In the case of the two-level bridge, we implement the same methodology with an additional step. First notice that $T$ can be approximated by $T~\approx~\sum_{k \in \{-1,1\}}~w^2(C E_k^{-1}~+~C^* E_k^{*\, -1})$  when $\lambda > \sigma$, such that
\begin{align}
  \frac{g(\varepsilon)}{(\varepsilon -i w) \pm v_{12}[x]} \approx \frac{g(\varepsilon)}{(\varepsilon - v_o -iw)  \mp v_o(x - \bar x)/\lambda}
\end{align}
holds. This assumption is valid for the system investigated in Figs.\ \ref{fig:tls} and \ref{fig:force}.\\
\vspace{0cm}

\nocite{*}
\bibliography{RefVoigt4}

\end{document}


\preprint{AIP/123-QED}

\title{Supplemental Information -- Generalized Voigt broadening due to thermal fluctuations of electromechanical nanosensors and molecular electronic junctions}

\author{Maicol A. Ochoa}
\affiliation{Biophysics Group, Microsystems and Nanotechnology Division, Physical Measurement Laboratory, National Institute of Standards and Technology, Gaithersburg, MD 20899}
\affiliation{Maryland Nanocenter, University of Maryland, College
Park, MD 20742}

\author{Michael Zwolak}
\email{mpz@nist.gov}
\affiliation{Biophysics Group, Microsystems and Nanotechnology Division, Physical Measurement Laboratory, National Institute of Standards and Technology, Gaithersburg, MD 20899}
\maketitle
\vspace{-1cm}
\section{The current $I (x)$.}
\vspace{-0.3cm}

\begin{center}
  \begin{figure}[b]
    \centering
    \includegraphics[scale=0.675]{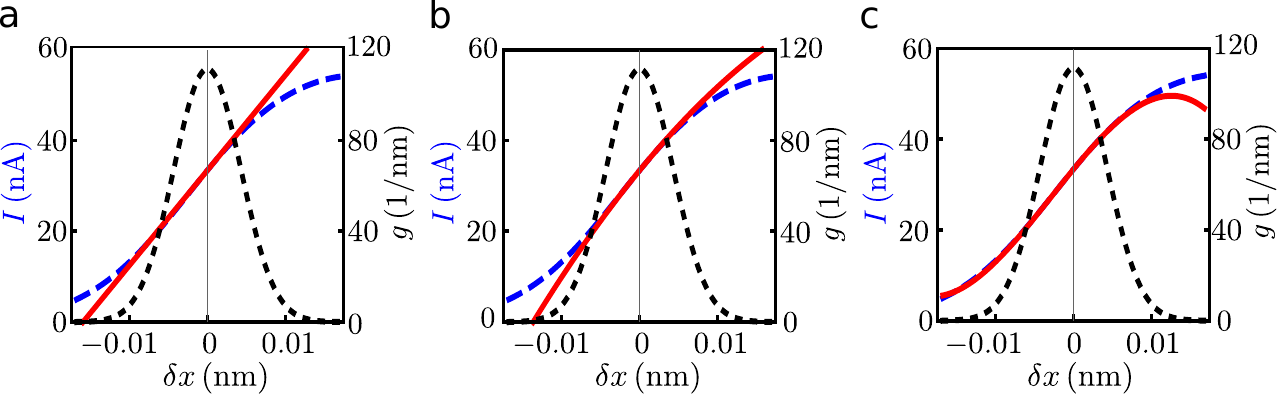}
    \caption{(Color online) Electronic current as function of the displacement $\delta x$ for the two-level system in Fig.\ 2. The numerically exact (blue, dashed) and approximate (red, solid) values for $I(x)$ with three different approximations. In (a), only the first order term in the expansion is considered, while, in (b) and (c), the approximation includes terms up to second and third order, respectively. The Gaussian probability distribution $g$ (black, dotted) shows the range of values accessible to the system through interactions with the environment at $300$~K. The bias is symmetrically applied about the Fermi level at $\mu=0.1$~eV. Other parameters for this system are as in Figs.\  2 and 3c.
    }
    \label{fig:currentX}
  \end{figure}
\end{center}

Here, we further investigate the approximate forms for the current and fluctuations. We start by considering the expression for the current $I(x)$, as a Taylor expansion near $x = \bar x$, i.e., near the equilibrium value. Up to third order

\begin{align}\label{eq:current_approx3}\tag{S1}
  I(x) \approx & I(\bar x) + \delta x\, \partial_{x} I(\bar x)+\frac{\delta x^2}{2} \partial_{x}^2 I(\bar x)+\frac{\delta x^3}{6} \partial_{x}^3 I(\bar x),
\end{align}
where $\delta x = x - \bar x$ is the displacement from $\bar x$. Figure~\ref{fig:currentX} shows the approximate and exact values of the current versus $\delta x$. Within the parameter region of interest, we notice that the first and second order approximations differ from the exact value for displacements that have a significant probability density. In order to correctly account for the contribution of these configurations at moderate $\delta x$, we must consider the approximate expansion up to third order.

\begin{center}
  \begin{figure}[H]
    \centering
    \includegraphics[scale=0.43]{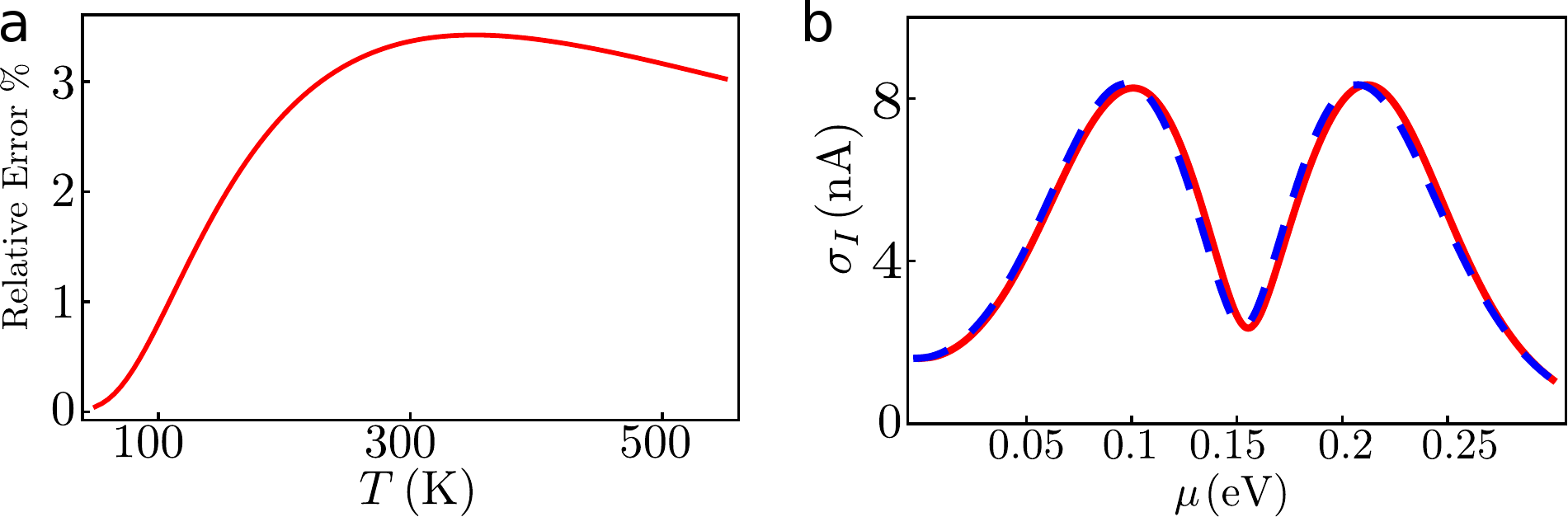}
    \vspace{-0.3cm}
    \caption{(Color online) Assessment of the approximate current and fluctuations. (a) Relative error, $(I_{\rm ex} -I_{\rm app})$, versus temperature for the approximation in Eq. \ref{eq:CurrAp}. The bias is applied symmetrically with Fermi level $\mu =0.12$~eV. (b) Current standard deviation as a function of the Fermi level $\mu$ for the two level system in Figs.\ 2 and 3, when undeflected (blue, dashed) and deflected (red, solid) by a force of $F = 100$~pN. Other parameters are as in Fig.\ 3. 
    }
    \label{fig:Idiff_var}
  \end{figure}
\end{center}

Next, we examine the behavior of the average current and the current fluctuations. From Eq.\ \eqref{eq:current_approx3} and considering Gaussian thermal fluctuations in $x$, such that $\langle \delta x \rangle = \langle \delta x^3 \rangle = 0 $, we find   
\begin{equation}\label{eq:CurrAp}\tag{S2}
  I_{\rm app} =\left . I (\bar x) + \frac{\sigma^2}{2} \partial_{x}^2 I \right|_{x = \bar x} \,
\end{equation}
with $\sigma^2 = (\beta \kappa)^{-1}$.  In Fig.\ \ref{fig:Idiff_var}, we observe that the difference between the exact thermally-averaged current $\langle I \rangle_{\rm ex}$ and the approximation in Eq. \eqref{eq:CurrAp}  increases with temperature up to a maximum value. This behavior is expected since higher temperatures increase the probability of large fluctuations. The decrease in the difference is observed, as a result of a decrease in the absolute current $\langle I \rangle$.  Figure \ref{fig:Idiff_var} also shows the current variance for the deflected and undeflected structure studied in Fig.\ 3c. This result indicates that the definition in Eq.\ (16) of the signal-to-noise ratio, in terms of the variance for the undeflected structure,  can also be given in terms of the current variance of the deflected structure with only minor differences. This holds only for systems undergoing weak forces or small local deflections.

\begin{center}
  \begin{figure}[H]
    \centering
    \includegraphics[scale=0.40]{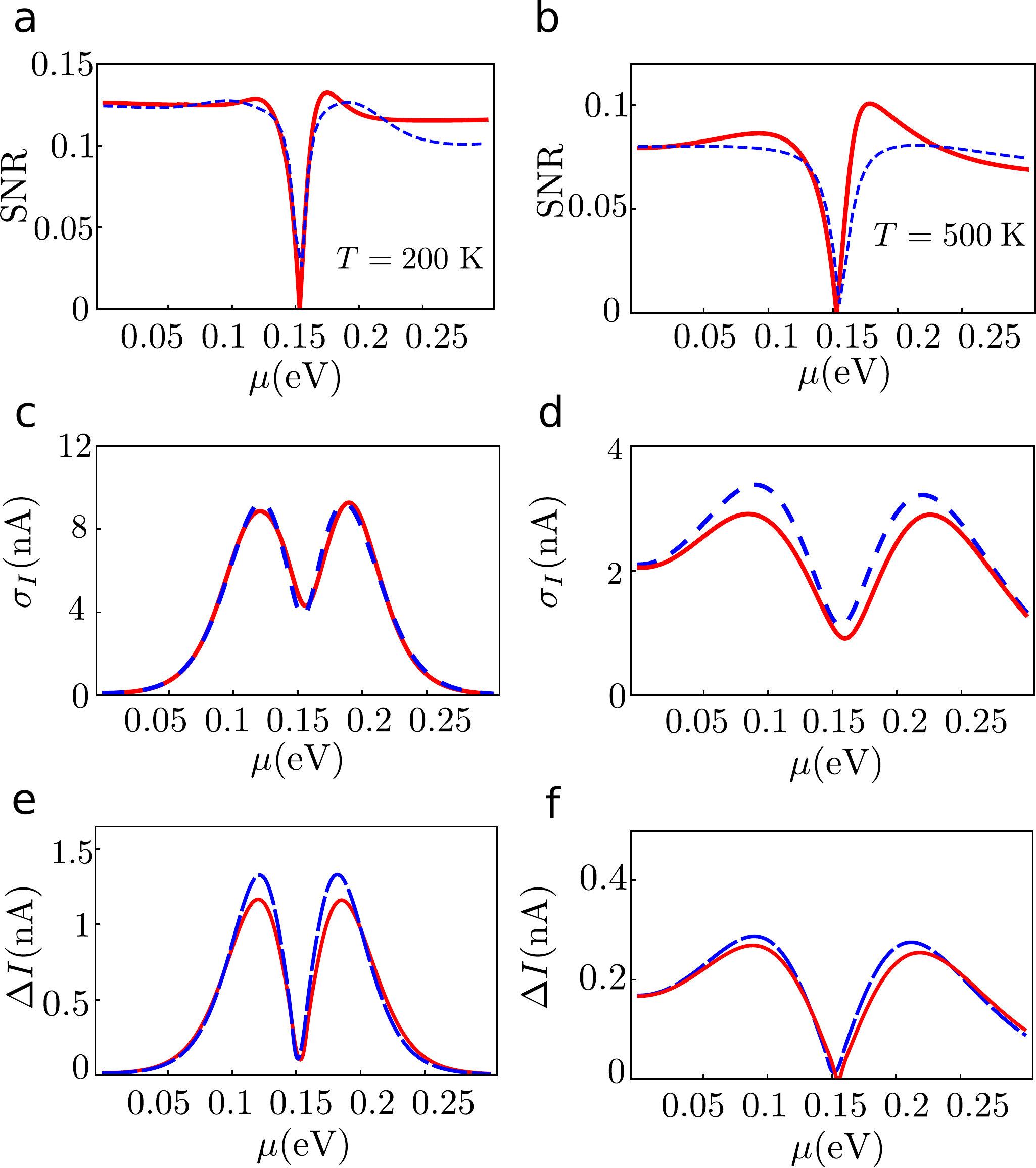}
       \vspace{-0.3cm}
    \caption{(Color online) Signal-to-noise ratio (SNR), $\sigma_I$ and $\Delta I$ for the two-level system and at two different temperatures. (a,b) SNR, (c,d) $\sigma_I$, and (e,f) $\Delta I$. Figures in (a,c,e) correspond to the system at 200 K, while in (b,d,f) it is 500 K. In every case, numerically exact (dashed, blue) and approximate calculations (red, solid) are shown. Other parameters are as in Fig.\ 3c.}
    \label{fig:FullComparison}
  \end{figure}
\end{center}

\begin{center}
  \begin{figure}[H]
    \centering
    \includegraphics[scale=0.41]{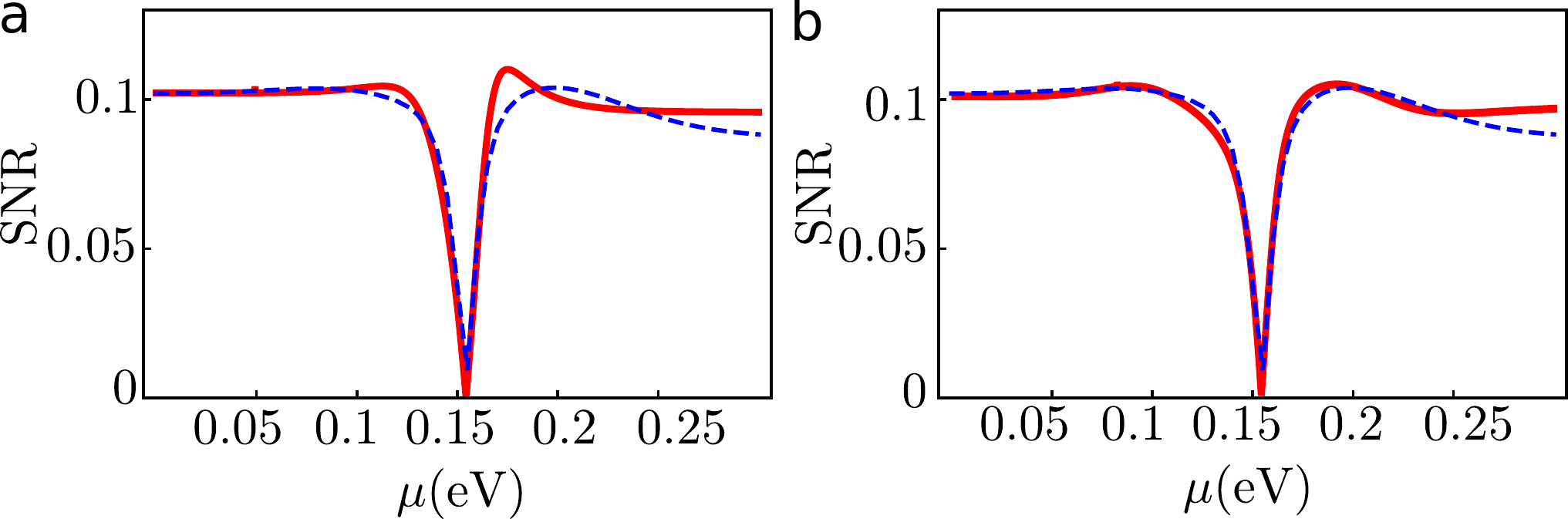}
       \vspace{-0.3cm}
    \caption{(Color online) SNR from two different approximations of the signal $\Delta I$. In (a), we adopt $\Delta I = \Delta x \, \partial_x I(\bar x)$ and, in (b),  $\Delta I = \langle I (F = 0) \rangle - \langle I (F = 100$ pN$) \rangle$, with $\langle I \rangle$ given by Eq.\ \eqref{eq:CurrAp}. Parameters for this model are the same as in Fig.\ 3d.
    }
    \label{fig:nonlin}
  \end{figure}
\end{center}

\vspace{-1cm}
We also find deviations of the approximate SNR expression from the exact values at high temperatures. In Figs. \ref{fig:FullComparison}a,b, we investigate this by considering the two-level system at two temperatures: 200~K and 500~K. These results suggest that the approximate form in Eq.~(22), is very accurate at low temperatures and close to the main feature (the depression at the molecular energy level) in the SNR. At high temperatures, the SNR is overestimated. Figure \ref{fig:FullComparison}d shows that this is because Eq.~(20) underestimates the magnitude of the current fluctuations in this regime. The local approximation to the signal $\Delta I$, in terms of the current derivative at $\bar x$, $\partial_x I$,  and the net displacement, $\Delta x$, is also more accurate at low temperatures (Figs.\ \ref{fig:FullComparison}e,f). However, even at moderate and low temperatures, we notice that this approximation leads to small differences far from the depression at the molecular level energy. In Figure \ref{fig:nonlin}, we show that this is due to the linear approximation of the $\Delta I$), by plotting the SNR at 300~K with $\Delta I$ calculated from Eq.\ \eqref{eq:CurrAp} (i.e., $ \langle I(0) \rangle- \langle I (F)\rangle$).  Consequently, better estimates of the SNR most account for higher order terms in the Taylor expansion of susceptibility (Eq.\ (18)).

\begin{center}
  \begin{figure}[htb]
    \centering
    \includegraphics[scale=0.44]{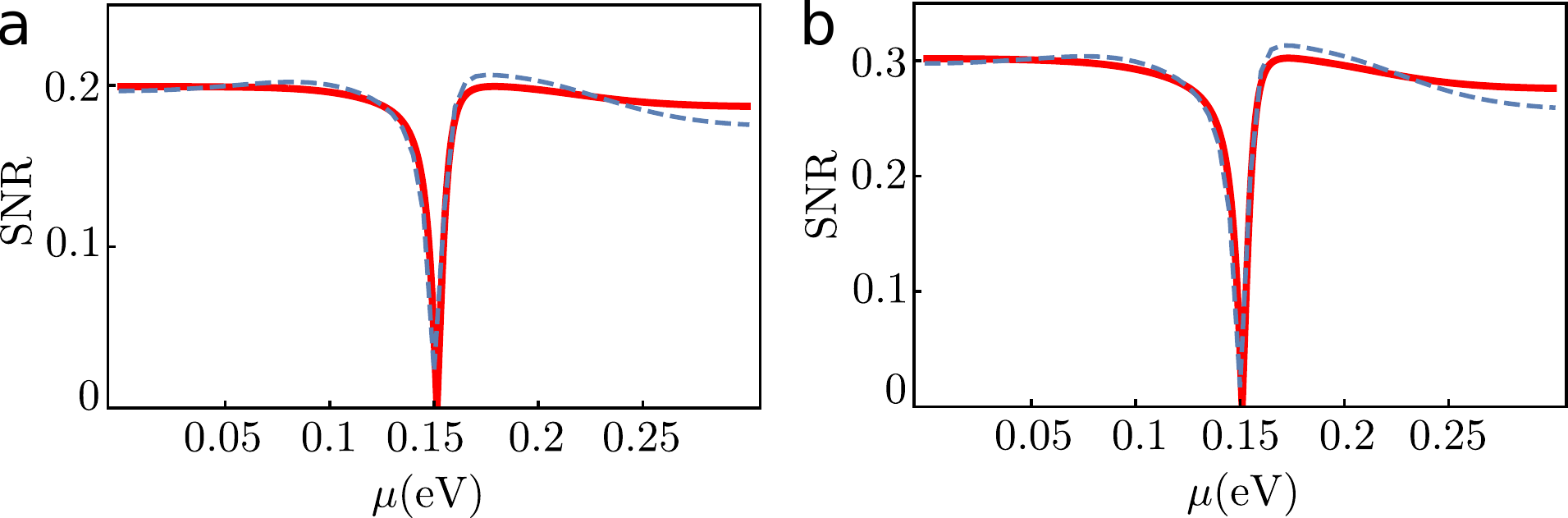}
    \caption{(Color online) SNR for the two-level system investigated in Fig. 3d, with modified spring constant $\kappa$. Numerically exact (blue, dashed) and approximate (red, solid) are shown. In (a) $\kappa = 5160$ eV/nm$^2$ and in (b) $\kappa = 3440$ ~eV/nm$^2$. Other paremeters are as in Fig.\ 3d.
    }
    \label{fig:kappa}
  \end{figure}
\end{center}

Last, in Fig. \ref{fig:kappa}, we show that the SNR, calculated at 300 K for the two-level system from Eq.\ (22), is still in good agreement with the exact results for more rigid, as well as more flexible, systems.


\vspace{-1.3cm}

\vspace{-1cm}

\vspace{-1.5cm}


\vspace{-1.5cm}
\section{Vibrational coupling}
In this section, we investigate the influence of vibrations on the Voigt profile and sensing protocols. We consider a junction consisting of a single level $\varepsilon_p$ coupled to a harmonic oscillator of frequency $\omega_o$, and Hamiltonian 
\begin{align}\tag{S3}
H(t) =& \omega_o \hat a^\dagger \hat a + \varepsilon_p \hat c^\dagger \hat c + M (\hat a^\dagger + \hat a) \hat c^\dagger \hat c,  
\end{align}
where $\hat a^\dagger $ ($\hat a$) is the vibrational creation (annhilation) operator, and $M$ is the coupling parameter. Utilizing the equation of motion (EOM), we find the EOM in the Keldysh contour for the electronic Green function $G(\tau_1, \tau_2) = -i \langle \hat c(\tau_1) \hat c^\dagger (\tau_2)\rangle$ in terms of the higher order correlation $-i \langle (\hat a + \hat a^\dagger)(\tau_1) \hat c(\tau_1) \hat c^\dagger (\tau_2)\rangle$.  

In order to compute the higher order correlation function, we adopt a semiclassical approximation in two steps. First, the retarded projection of this correlation function, in real time, is factorized in terms of lower order correlations functions as 

\begin{align}
  -i \langle (\hat a + \hat a^\dagger)(t_1) &\hat c(t_1) \hat c^\dagger (t_2)\rangle \notag\\
                                                   \approx & \int dt  \langle (\hat a + \hat a^\dagger)(t_1- t)\rangle G^r(t, t_2). \tag{S4}
\end{align}
This retains a memory of a vibrational interaction with a transporting electron.
Next, the expectation $ \langle (\hat a + \hat a^\dagger) (t) \rangle $ is calculated starting from the Langevin equation for a dissipative harmonic oscillator, such that
\begin{align}\tag{S5}
  \langle (\hat a^\dagger + \hat a) (t) \rangle& \approx \sqrt{\frac{2 m \omega_o}{\hbar}} x(t),\label{eq:Ap_x_osc}
\end{align}
together with
\begin{align}\label{eq:Ap_EOM_osc}\tag{S6}
  \frac{d^2}{dt^2} (x(t) - \bar x)  =& \omega_o^2 (x(t) - \bar x)- \eta \frac{d}{dt } x(t) + \frac{F(t)}{m}.
\end{align}


\begin{center}
  \begin{figure}[htb]
    \centering
    \includegraphics[scale=0.64]{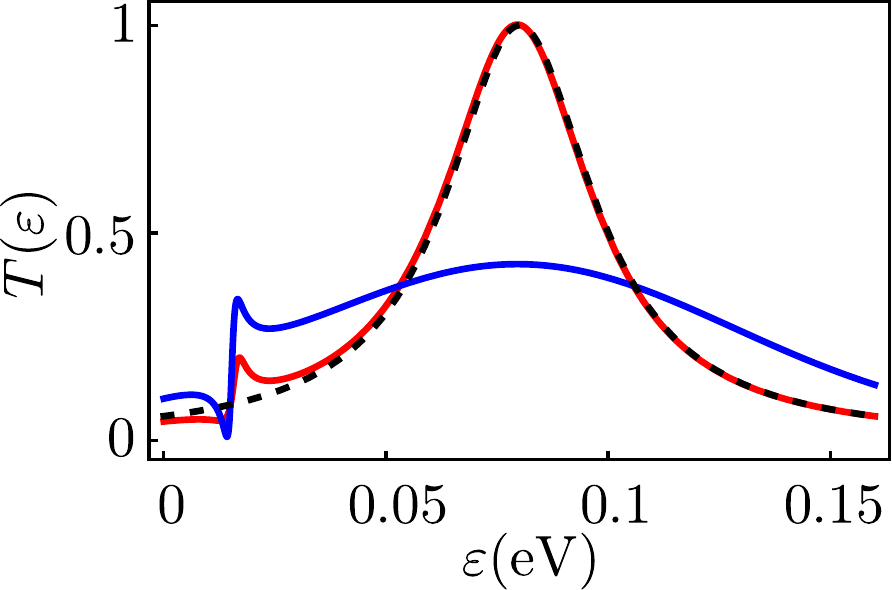} 
    \caption{(Color online)
    Transmission function for a single level coupled to a harmonic oscillator with (blue, solid) and without energy-level fluctuations (red, solid). We also show the $T(\varepsilon)$ for the uncoupled energy level (black, dashed) at $\bar \varepsilon_p$. The vibration gives an additional peak in the transmission function close to the harmonic oscillator's characteristic frequency. We asume a white noise, with $\mathcal{R}(\omega) = 1$ eV/m . Parameters for this model are $\bar \varepsilon_p =~80$~meV, $\omega_o = 15$~meV, $m = 12$~amu, $M = 0.015$~meV, $\hbar \eta = $ 4~meV, $\sigma = 42$~meV. The frequency $\omega_o$ in this model is larger than the relevant vibrational modes for graphene deflectometry.}
    \label{fig:VoigtInteracting}
  \end{figure}
\end{center}


In Eqs.\ \eqref{eq:Ap_x_osc} and \eqref{eq:Ap_EOM_osc}, $m$ is the mass of the oscillator,  $\eta$ is the friction coefficient originating on several relaxations mechanisms, and $R(t)$ is a random force due to environmental fluctuations. The Fourier transform of $x(t)- \bar x$ is therefore
\begin{align}\tag{S7}
  \mathcal{X} (\varepsilon) &= \frac{\mathcal{F}(\varepsilon/\hbar)/m}{\omega_o^2 - (\varepsilon/\hbar)^2 - i \eta \varepsilon/\hbar},
\end{align}
leading to the retarded electronic Green Function $G^{\rm r}(\varepsilon)$ that effectively accounts for the vibrational coupling,
\begin{align}\tag{S8}
  G^r(\varepsilon) &= \frac{1}{\varepsilon - (\varepsilon_p + M' {\rm Re}\mathcal{X}(\varepsilon)) +i (w - M' {\rm Im } \mathcal{X}(\varepsilon))},
\end{align}
with $M' = M \sqrt{ 2 m \omega_o/\hbar}$. Consequently,
\begin{align}\tag{S9}
  T(\varepsilon, \varepsilon_p) =& \frac{w^2}{(\varepsilon -\varepsilon_p- M' {\rm Re} \mathcal{X})^2 + (w - M' {\rm Im} \mathcal{X})^2}. \label{eq:apTransM}
\end{align}
Notice that in the abscence of coupling ($M=0$), the electronic transmission function in Eq.\ ~\eqref{eq:apTransM} coincides with the form derived in Eq.\ (2). In the weak electron-phonon coupling regime ($M \ll w$), and when environmental fluctuations induce broadening to the energy level $\varepsilon_p$, in the form described in Eqs.\ (3)-(5), the thermally broadened electronic transmission function assumes the same functional form than Eq.\ (6) with $E$ replaced by $E' = i (\varepsilon - M' {\rm Re} \mathcal{X} - \bar \varepsilon_p) + w - M' {\rm Im} \mathcal{X}$. We notice that by writing the transmission function in Eq.\ \eqref{eq:apTransM} as a truncated Taylor expansion to first order, we can decompose the current in the form $I \approx I_{o}+ I_{\rm vib}$, with
\begin{align}\tag{S10}
  I_{\rm vib} \sim M' \int \frac{d \varepsilon}{2 \pi} T(\varepsilon)^2 {\rm Im}\mathcal{X}.
\end{align}
Figure \ref{fig:VoigtInteracting} shows the Voigt profile for the electronic transmission function of a single level weakly coupled to the local vibration. While an additional channel to transport is imprinted in the Voigt profile, at room temperature, the net contribution of these currents may not notably affect the general profile of the current under bias (see Fig.\ \ref{fig:VoigtInteracting}).   Also, as described in Ref.\ 37, the linewidth of the spectral peak in inelastic spectra carry intrinsic smearing with a width at half maximum proportional to  $1/\beta$, limiting its detectability at room temperature. A similiar limiting factor will occur here.


\begin{center}
  \begin{figure}[htb]
    \centering
    \includegraphics[scale=0.4]{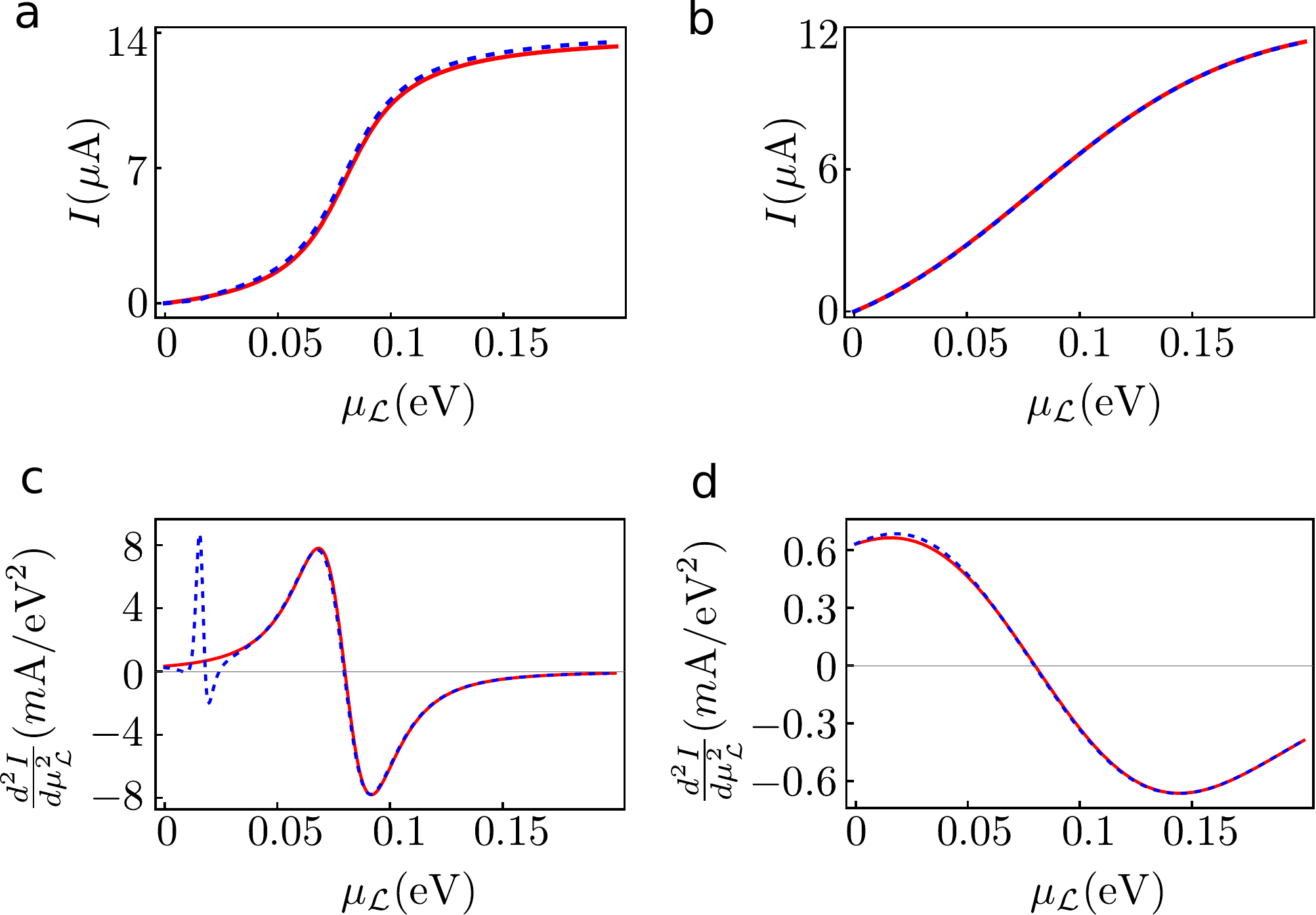} 
    \caption{(Color online) Current (a,b) and associated inelastic spectrum (c,d) for a noninteracting single level (red, solid) and coupled to a local vibration (blue, dashed) with coupling strength $M=0.015$ meV. Bias $\mu_{\mathcal{L}}$ is applied to the left contact with $\mu_{\mathcal{R}}=0$. The results in (a,c) where obtained at $10$ K and assuming no fluctuations in the level energy $\varepsilon_{\rm p}$, while (b,d) are calculated at $300$~K  with an energy variance $\sigma = 42$~meV.  Other parameters for this system the same as in Fig.\ \ref{fig:VoigtInteracting}. This result illustrates that vibrationally-assisted currents are most significant at lower temperatures. See also Ref.\ ~37.}
    \label{fig:VoigtInteracting}
  \end{figure}
\end{center}
